\documentclass[aps,prl,twocolumn,superscriptaddress,showpacs]{revtex4}

\usepackage{graphicx}

\begin{document}

\def\lufesi{Lu$_{2}$Fe$_{3}$Si$_{5}$}
\def\tc{$T_{c}$}
\def\rfesi{$R_{2}$Fe$_{3}$Si$_{5}$}
\def\scfesi{Sc$_{2}$Fe$_{3}$Si$_{5}$}
\def\yfesi{Y$_{2}$Fe$_{3}$Si$_{5}$}
\def\tmfesi{Tm$_{2}$Fe$_{3}$Si$_{5}$}
\def\erfesi{Er$_{2}$Fe$_{3}$Si$_{5}$}


\title{Specific heat evidence for two-gap superconductivity in ternary-iron silicide {\lufesi}}


\author{Y.~Nakajima}
\author{T.~Nakagawa}
\author{T.~Tamegai}
\affiliation{Department of Applied Physics, The University of Tokyo, Hongo, Bunkyo-ku, Tokyo 113-8656, Japan}

\author{H.~Harima}
\affiliation{Department of Physics, Kobe University, Kobe 657-8501, Japan}


\date{\today}

\begin{abstract}
We report low-temperature specific heat studies on single-crystalline ternary-iron silicide superconductor {\lufesi} with {\tc} = 6.1 K down to $\sim${\tc}$/20$. We confirm a reduced normalized jump in specific heat at {\tc}, and find that the specific heat divided by temperature $C/T$ shows sudden drop at $\sim${\tc}$/5$ and goes to zero with further decreasing temperature. These results indicate the presence of two distinct superconducting gaps in {\lufesi}, similar to a typical two-gap superconductor MgB$_{2}$. We also report Hall coefficients, band structure calculation, and the anisotropy of upper critical fields for {\lufesi}, which support the anisotropic multiband nature and reinforce the existence of two superconducting gaps in {\lufesi}. 
\end{abstract}

\pacs{74.25.Bt, 74.70.Dd, 71.18.+y}

\maketitle



Since the discovery of high-{\tc} superconducting intermetallic compound MgB$_{2}$ with $T_{c}\sim$ 39 K \cite{nagamatsu}, the materials containing light elements, such as, B, C, and Si, has attracted much attention. Among compounds containing light elements, the ternary-iron silicide superconductors {\rfesi} with $R$ = Lu, Y, Sc, Tm, or Er, have stimulated many investigations of the relation between superconductivity and magnetism due to magnetic atoms \cite{braun,segre2,noguchi}. {\rfesi} crystallizes in the tetragonal {\scfesi}-type structure consisting of quasi-one-dimensional iron chain along $c$-axis and quasi-two-dimensional iron square structures \cite{chabot}. The properties of these compounds are very peculiar in several aspects. For instance, one of the highest superconducting transition temperature among any known iron compounds are observed in {\lufesi} with {\tc} = 6.1 K and {\scfesi} with {\tc} = 4.5 K \cite{braun, shirotani}. In addition, a large negative pressure effect on {\tc} is reported in {\lufesi} and {\scfesi} ($dT_{c}/dp=-7\times 10^{-5}$ K/ bar) while a large enhancement of {\tc} is exhibited in {\yfesi} ($T_{c}/dp=-33\times 10^{-5}$ K/ bar) \cite{segre}. Moreover, reentrant superconductivity is found for {\tmfesi} \cite{segre2} and the coexistence between superconductivity and antiferromagnetism is observed in {\erfesi} \cite{noguchi}.

One of the most puzzling issues in the ternary-iron silicide superconductors is the peculiar superconducting state in {\lufesi}. In this material, significant non-BCS features, such as a large residual linear term in the superconducting specific heat and a reduced normalized specific heat jump at {\tc} smaller than BCS value of 1.43, have been reported \cite{vining,stewart,tamegai}. In addition, a non-exponential power-law specific heat below {\tc} has been observed \cite{vining,stewart,tamegai}. Moreover, a rapid decrease in {\tc} when a small amount of nonmagnetic impurities replace some of Fe atoms \cite{xu}. These anomalous features let us speculate a possibility of the spin-triplet superconductivity in {\lufesi}. However, the Josephson effect measurement between {\lufesi} and Nb is against the triplet pairing state \cite{noer}. It has been an unsettled problem whether unusual superconducting properties of {\lufesi} are intrinsic  or arise from magnetic impurities due to structural disorder and defects. In this letter, we address the issue of the unusual superconducting properties of {\lufesi} by the low temperature specific heat measurements of high-quality single crystals. We report that the peculiar temperature dependence of specific heat can be well-described by the phenomenological two-gap model similar to a typical two-gap superconductor MgB$_{2}$. We also investigate the Hall effect, band structure, and the anisotropy of upper critical fields for {\lufesi}, in order to support the validity of two-gap model in terms of electronic structure.

Single crystals are grown by the floating-zone technique using an image furnace  in Ar atmosphere at a rate of $\sim$2 mm/h. The starting polycrystalline rod is prepared by fusing several polycrystalline lumps, prepared by arc melting the constituent elements in Ar atmosphere. The single crystals are annealed for 4 weeks at 800$^{\circ}$C followed by 5 days at 1250$^{\circ}$C. It should be noted that the annealing at high temperature for an extended period of time is essential to obtain sharp superconducting transition. Magnetization was measured by a commercial SQUID magnetometer (MPMS-XL5, Quantum Design). The Hall coefficient  measurements are performed by standard six-wire configuration. Specific heat was measured by the relaxation method in a $^{3}$He refrigerator. A band structure calculation is carried out by using a full potential APW (FLAPW) method with the local density approximation (LDA) for the exchange correlation potential.

\begin{figure}
\includegraphics[width=8cm]{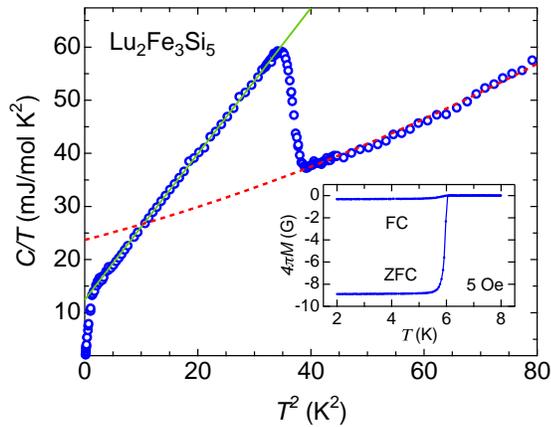}
\caption{(color online) The specific heat divided by temperature $C/T$ as a function of $T^{2}$ for {\lufesi}. The dashed line represent the fit to the data in the normal state $C=\gamma_{n}T+\beta_{n}T^{3}+\alpha_{n}T^{5}$. The solid line is the fit to the data in the superconducting state $C=\gamma_{s}T+\beta_{s}T^{3}$. Inset shows temperature dependence of the zero-field-cooled (ZFC) and field-cooled (FC) magnetization for {\lufesi}  at 5 Oe. \label{FIG1}}
\end{figure}

Figure \ref{FIG1} depicts the specific heat divided by temperature $C/T$ as a function of $T^{2}$ for {\lufesi}. A jump in the specific heat at 6.1 K indicates that the transition into the superconducting state is very sharp, which is also observed in magnetization measurements shown in the inset of Fig. \ref{FIG1}. Above {\tc}, $C/T$ slightly deviates from $T^{2}$-dependence. The normal-state specific heat, which keeps the entropy balance at {\tc}, is fitted by using the equation, $C=\gamma_{n}T+\beta_{n}T^{3}+\alpha_{n}T^{5}$, where $\gamma_{n}T$ is the electronic term and $\beta_{n}T^{3}+\alpha_{n}T^{5}$ represents the phonon contribution. We obtain the parameters $\gamma_{n}$ = 23.7 mJ/mol K$^{2}$,  $\beta_{n}$ = 0.276 mJ/mol K$^{4}$, and $\alpha_{n}$ = 1.73$\times 10^{-3}$ mJ/mol K$^{6}$, which are very close to the values for the previous polycrystalline samples \cite{vining, stewart, tamegai}. The large $T^{5}$-term observed in the normal-state specific heat suggests complex phonon density of states and has been reported for other ternary superconductors \cite{woolf}. The magnitude of the jump at {\tc}, $\Delta C$, is 150 mJ/mol K. The value of the normalized specific heat jump at {\tc},  $\Delta C/\gamma_{n}T_{c}$, is 1.05, significantly smaller than the BCS value of 1.43, close to reported values \cite{vining, stewart, tamegai}. Below {\tc} the specific heat for {\lufesi} shows peculiar $T$-dependence. At high temperatures from $\sim T_{c}$ to $\sim T_{c}/5$, $C/T$ shows almost perfectly $T^{2}$-dependence. We estimated a linear extrapolation term, $\gamma_{s}$, from {\tc} to $T=0$ by using the equation, $C=\gamma_{s}T+\beta_{s}T^{3}$. From the fit to the data plotted in Fig. \ref{FIG1}, we obtain the parameters $\gamma_{s}$ = 12.5 mJ/mol K$^{2}$ and $\beta_{s}$ = 1.37  mJ/mol K$^{4}$. The linear term $\gamma_{s}$ is about 53 \% of the normal-state electronic coefficient. These values are again consistent with those reported previously \cite{vining, stewart, tamegai}.  The observation of the reduced normalized jump at {\tc} and the large linear extrapolation term in the superconducting specific heat for both polycrystalline and high-quality single crystalline samples strongly suggest that these anomalous features stems from intrinsic origin.

\begin{figure}
\includegraphics[width=8cm]{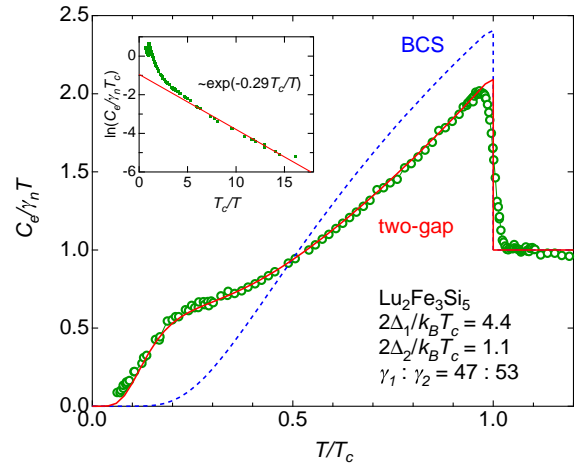}
\caption{(color online) Temperature dependence of the normalized electronic specific heat for {\lufesi}. Dashed and solid lines are BCS normalized specific heat and a two-gap fit calculated by Eqs. (\ref{eq1}) and (\ref{eq2}), respectively. Inset shows semi-logarithmic plot of $C_{e}/\gamma_{n}T_{c}$ vs $T_{c}/T$ for {\lufesi}. The solid line is a linear fit below $\sim T_{c}/5$. \label{FIG2}}
\end{figure}

We now discuss the detailed temperature dependence of specific heat for {\lufesi}. Figure \ref{FIG2} shows temperature dependence of the normalized electronic specific heat for {\lufesi}. For comparison, we also plot the BCS weak-coupling limit specific heat (dashed line) calculated based on the following equations \cite{bouquet2}, 
\begin{equation}
	\frac{S}{\gamma_{n}T_{c}}=-\frac{6}{\pi^{2}}\frac{\Delta_{0}}{k_{B}T_{c}}\int^{\infty}_{0}[f\ln f+(1-f)\ln(1-f)]dy,\label{eq1}
\end{equation}
\begin{equation}
	\frac{C}{\gamma_{n}T_{c}}=t\frac{d(S/\gamma_{n}T_{c})}{dt},\label{eq2}
\end{equation}
where $f=(\exp(\beta E)+1)^{-1}$ and $\beta=(k_{B}T)^{-1}$. The energy of quasi-particles is given by $E=(\epsilon^{2}+\Delta^{2}(t))^{1/2}$, where $\epsilon$ is the energy of the normal electrons relative to the Fermi energy, and the temperature dependence of an energy gap varies as  $\Delta(t)=\Delta_{0}\delta(t)$, where  $\delta(t)$ is the normalized BCS gap at the reduced temperature $t=T/T_{c}$ as tabulated in Ref.\cite{muhlschlegel}. The integration variable is $y=\epsilon/\Delta_{0}$. Apparently, temperature dependence of $C_{e}/\gamma_{n}T$ for {\lufesi} is very different from that of BCS specific heat. In the BCS weak-coupling limit, below {\tc}, $C_{e}/\gamma_{n}T$ drastically decreases with decreasing temperature and is almost zero below {\tc}/5. In contrast to the BCS weak-coupling  limit,  $C_{e}/\gamma_{n}T$ for {\lufesi} decreases slowly with decreasing temperature from {\tc} to $\sim$ {\tc}/5. $C_{e}$ remains about 60 \% of normal-state specific heat even at $\sim$ {\tc}/5. Below $\sim${\tc}/5,  $C_{e}/\gamma_{n}T$ shows a sudden second drop and goes to zero with further decreasing temperature. Very similar behavior is also observed in MgB$_{2}$ \cite{bouquet,bouquet2,yang,wang}. In order to explain the anomalous behavior of specific heat for MgB$_{2}$, two-gap model has been proposed and it reproduces the peculiar specific heat well \cite{bouquet2,bouquet}. We show here similar analysis for {\lufesi}. Within the two-gap model, the total specific heat can be given by the sum of the contribution from each gap calculated independently using the Eqs. (\ref{eq1}) and (\ref{eq2}). The solid line shown in Fig. \ref{FIG2} represents the two-gap fit with the magnitude of two distinct superconducting gaps, $2\Delta_{1}/k_{B}T_{c}=4.4$ and $2\Delta_{2}/k_{B}T_{c}=1.1$, and the ratio of weight, $\gamma_{1}:\gamma_{2}=47:53$, where a partial electronic specific heat coefficient $\gamma_{i}$ with $\gamma_{n}=\gamma_{1}+\gamma_{2}$ characterizes each band. The calculation based on the two-gap model shown in Fig. \ref{FIG2} well reproduces the experimental results, supporting the presence of distinct two superconducting gaps in {\lufesi} similar to MgB$_{2}$. At high temperatures just below {\tc}, the magnitude of $\Delta C/\gamma_{n}T$ is determined by $\gamma_{1}$ because quasi-particles excited across the smaller gap $\Delta_{2}$ have negligible contribution  to the specific heat jump at {\tc}. By contrast, at low temperatures $T\ll T_{c}$, $C/T$ shows the sudden decrease and its magnitude is determined by $\gamma_{2}$ because the smaller gap $\Delta_{2}$ determines the excitation of quasi-particles. Therefore, all the anomalous features observed in the superconducting specific heat for {\lufesi}, such as the reduced jump of the specific heat at {\tc}, the large residual linear term, and the sudden decrease below $\sim T_{c}/5$, can be well described by the two-gap model. Similar results are obtained from other single crystalline samples. It should be noted that penetration depth measurements also support the two-gap superconductivity in {\lufesi} \cite{gordon}.

\begin{figure}
\includegraphics[width=8cm]{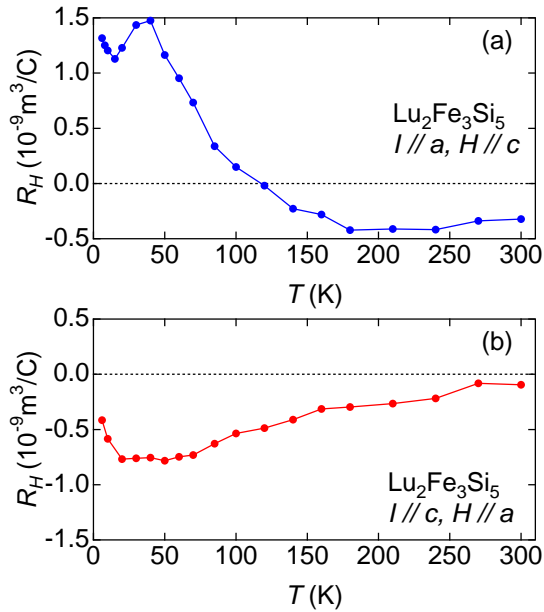}
\caption{(color online) Temperature dependence of the Hall coefficients for {\lufesi} with (a)$I\parallel a, H\parallel c$ and (b)$I\parallel c, H\parallel a.$\label{FIG3}}
\end{figure}

A different approach reinforces the existence of smaller superconducting gap. In the weak-coupling BCS superconductors, it is well known that empirical relation $C_{e}\propto\exp(-aT_{c}/T)$ with $a=1.44$ holds for $2.5<T_{c}/T<6$ \cite{gladstone}. The inset of Fig. \ref{FIG2} shows a semi-logarithmic plot of $C_{e}/\gamma_{n}T_{c}$ vs $T_{c}/T$. Solid line represents a linear fit to the data at low temperatures following the relation $C_{e}/\gamma_{n}T_{c}\sim\exp(-aT_{c}/T)$ with $a=0.29$. Obviously, the observed value of $a$ is about four times smaller, suggesting that a fully opened gap exists at low temperatures but its magnitude is very small. All these features are observed in MgB$_{2}$ \cite{yang, bouquet}.


\begin{figure}
\includegraphics[width=6cm]{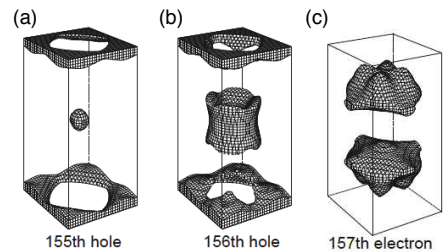}
\caption{Fermi surfaces of {\lufesi} calculated by a FLAPW method: hole-like Fermi surfaces from (a) 155th and (b) 156th bands, and (c) electron-like Fermi surface from 157th band. \label{FIG4}}
\end{figure}

The temperature dependence of Hall coefficients also supports that {\lufesi} is a multiband system. Figures \ref{FIG3}(a) and (b) depict the temperature dependence of Hall coefficients $R_{H}$ for {\lufesi}. In the configuration with $I\parallel a$ and $H\parallel c$, at high temperatures near 300 K, the sign of $R_{H}$ is negative. $R_{H}$ begins to increase with decreasing temperature from $\sim$150 K. At $\sim$ 110 K, the sign of $R_{H}$ changes to positive. With further decreasing temperature, $R_{H}$ decreases after showing a maximum at  around 40 K.  At low temperatures, $R_{H}$ increases again with decreasing temperature after showing a minimum. In the configuration with $I\parallel c$ and $H\parallel a$, the sign of $R_{H}$ is negative in the whole temperature range. At high temperatures, $R_{H}$ decreases with decreasing temperature and the slop becomes slightly larger below $\sim$ 150 K. $R_{H}$ becomes nearly constant below 70 K and increases steeply with decreasing temperature below 20 K. In a simple metal, the Hall coefficient in single band system is written as, $R_{H}=\pm 1/ne$, where $n$ is a carrier number. $R_{H}$ is almost $T$-independent. In contrast, the Hall coefficient in multiband system, for instance, consisting of electron and hole bands, is given by, $R_{H}= (n_{h}\mu_{h}^{2}-n_{e}\mu_{e}^{2})/(e(n_{h}\mu_{h}+n_{e}\mu_{e})^{2})$, where $n_{h}$ ($n_{e}$) is a carrier number of hole (electron) band and $\mu_{h}$ ($\mu_{e}$) is a mobility of hole (electron) band. The Hall coefficient in multiband system can be temperature dependent and even show the sign change because of the different temperature dependence of the mobility in each band. The strong temperature dependence of the Hall coefficient for {\lufesi} in both configurations strongly suggests that {\lufesi} is a multiband system.

The Fermi surfaces in MgB$_{2}$, where two-gap superconductivity has been studied extensively, is formed by two kinds of bands: two-dimensional $\sigma$-band of B and three-dimensional $\pi$-band of B \cite{kortus, harima,tsuda}. It is believed that the larger gap opens in $\sigma$-band while the smaller gap opens in $\pi$-band. Figure \ref{FIG4} shows the Fermi surfaces for {\lufesi} calculated by an FLAPW method. The Fermi surfaces consist of two hole-like bands (155th and 156th) and one electron-like band (157th), each having contribution to the density of states at the Fermi level $N(0)$, 13.9 \%, 44.5 \%, and 41.6 \%, respectively. In the two hole-like bands, some part of Fermi surfaces are quasi-one-dimensional, reflecting the iron-chain structure, while others are three dimensional. Unfortunately, it is not clear which parts of the bands shown in Fig.\ref{FIG4} are responsible for the larger and the smaller gaps. From the analogy of MgB$_{2}$, however, we speculate that the existence of anisotropic band and three-dimensional one may play an important role in the two-gap superconductivity. Evaluation of Fermi surface dependent gap as has been done for MgB$_{2}$ is necessary to clarify this point \cite{choi}. From $N(0)$ obtained by the band calculation,  we can obtain the calculated specific heat coefficient $\gamma_{band}$ = 8.69 mJ/mol K$^{2}$, which is about three times smaller than $\gamma_{n}$ in the present study, indicating that presence of the strong electron-phonon and/or electron-electron interactions.

\begin{figure}
\includegraphics[width=8cm]{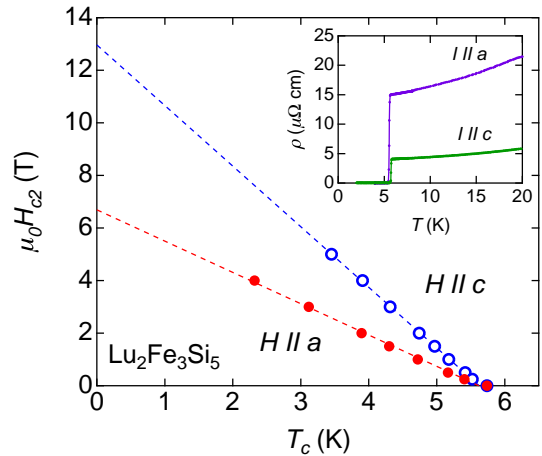}
\caption{(color online) Temperature dependence of upper critical field $\mu_{0}H_{c2}$ along $a$-axis ($\bullet$) and $c$-axis ($\circ$) obtained by resistivity measurements. Dashed lines are linear fits to the data. Inset shows the temperature dependence of zero-field inter- ($I\parallel c$) and in-plane ($I\parallel a$) resistivity. \label{FIG5}}
\end{figure}

Figure \ref{FIG5} shows temperature dependence of upper critical field along $a$-axis and $c$-axis obtained by resistivity measurements. The temperature dependences of $H_{c2}$ in both directions are almost linear in the present temperature region. Apparently, as shown in Fig. \ref{FIG5}, $H_{c2}^{c}$ is larger than $H_{c2}^{a}$ in the whole temperature range. Anisotropy of upper critical field $H_{c2}^{c}/H_{c2}^{a}$ is $\sim$ 2, indicating that {\lufesi} is a weakly one-dimensional superconductor reflecting the crystal structure. In fact, the zero-field inter-plane resistivity ($I\parallel c$) is about four times smaller than in-plane resistivity ($I\parallel a$) just above $T_{c}$ as shown in the inset of Fig. \ref{FIG5}.

In summary, we have prepared high-quality single crystals of ternary iron-silicide superconductor {\lufesi} and performed the low-temperature specific heat studies down to $\sim T_{c}/20$. The measurements reveal a reduced normalized specific heat jump at {\tc}, a large linear term even at $\sim T_{c}/5$, and a second drop of $C_{e}/T$ below {\tc}/5. These results can be well reproduced by the phenomenological two-gap model similar to a typical two-gap superconductor MgB$_{2}$. The Hall coefficient measurements and band structure calculations support the multi-gap superconductivity. We emphasize that {\lufesi} gives a unique opportunity to investigate the multi-gap superconducting properties in the following reasons: (1) simplicity of the multi-gap nature, {\it i.e.}, each band has nearly the same contribution to density of states, (2) relatively high transition temperature, and (3) large crystal size. These advantage, as compared with several other possible multigap superconductors, such as MgB$_{2}$, NbSe$_{2}$ \cite{guritanu}, PrOs$_{4}$Sb$_{12}$ \cite{seyfarth}, URu$_{2}$Si$_{2}$ \cite{kasahara} may shed light on the details of multi-gap superconductivity.

We thank A. Koshelev, R. Prozorov, and K. Kumagai for stimulating discussions.  This work was partly supported by a Grant-in-Aid for Scientific Reserch from the Ministry of Education, Culture, Sports, Science and Technology.


\begin{thebibliography}{99}
\bibitem{nagamatsu} J. Nagamatsu, N. Nakagawa, T. Muranaka, Y. Zenitani, J. Akimitsu, Nature (London) \textbf{410}, 63 (2001).
\bibitem{braun}  H. Braun, Phys. Lett. \textbf{75A}, 386 (1980).
\bibitem{segre2} C.U. Segre and H. F. Braun, Phys. Lett. \textbf{85A}, 372 (1981).
\bibitem{noguchi}S. Noguchi and K. Okuda, Physica B \textbf{194-196}, 1975 (1994).
\bibitem{chabot} B. Chabot, and E. Parthe, J. Less-Common Metals \textbf{97}, 285 (1984).
\bibitem{shirotani}I. Shirotani, Y. Shimaya, K.Kihou, C. Sekine, N. Takeda, M. Ishikawa and T. Yagi, J. Phys.: Condens. Matter \textbf{15}, S2201 (2003).
\bibitem{segre} C. U. Segre and H. F. Braun, {\it Physics of Solids Under High Pressure}, edited by J. S. Schilling and R. N. Sheloton (North-Holland, Amsterdam, p. 381 (1981).
\bibitem{vining} C. B. Vining, R. N. Shelton, H. F. Braun, and M. Pelizzone, Phys. Rev. B \textbf{27} ,2800 (1983) .
\bibitem{stewart} G. R. Stewart, G. P. Meisner, and C. U. Segre, J. Low Temp. Phys. \textbf{59}, 237  (1985).
\bibitem{tamegai} T. Tamegai, T. Nakagawa, and M. Tokunaga, Physica C \textbf{460-462} (2007) 708.
\bibitem{xu}Y. Xu and R. N. Shelton, Solid State Commun. \textbf{68}, 395 (1988).
\bibitem{noer}R. J. Noer, T. P. Chen, and E. L. Wolf, Phys. Rev. B \textbf{31} (1985) 647.
\bibitem{woolf} L. D. Woolf, D. C. Johnston, H. B. Mackay, R. W. MacCallum, and B. Maple, J. Low Temp. Phys. \textbf{35}, 651 (1979).
\bibitem{bouquet2} F. Bouquet. Y. Wang, R. A. Fisher, D. G. Hinks, J. D. Jorgensen, A. Junod, and N. E. Phillips, Europhys. Lett. \textbf{56}, 856 (2001).
\bibitem{muhlschlegel}B. M\"uhlschlegel, Z. Phys. \textbf{155}, 313 (1959).
\bibitem{bouquet}F. Bouquet, R. A. Fisher, N. E. Phillips, D. G. Hinks, and J. D. Jorgensen, Phys. Rev. Lett. \textbf{87}, 047001 (2001).
\bibitem{yang}H. D. Yang, J. -Y. Lin, H. H. Li, F. H. Hsu, C. J. Liu, S. -C. Li, R. -C. Yu and C. -Q. Jin, Phys. Rev. Lett. \textbf{87}, 167003 (2001).
\bibitem{wang}Y. Wang, T. Plackowski, and A. Junod, Physica C \textbf{355}, 179 (2001).
\bibitem{gordon} R. Gordon, M. D. Vannette, C. Martin, Y. Nakajima, T. Tamegai, and R. Prozorov, to be submitted.
\bibitem{gladstone}G. Gladstone, M. A. Jensen, and J. R. Schrieffer, in {\it Superconductivity}, edited by R. D. Parks (Marcel Dekker, New York, 1969), Vol. 2.
\bibitem{kortus}J. Kortus, I.I. Mazin, K. D. Belashchenko, V. P. Antropov, and L. L. Boyer, Phys. Rev. Lett. \textbf{86} 4656 (2001).
\bibitem{harima}H. Harima, Physica C \textbf{378-381}, 18 (2002).
\bibitem{tsuda}S. Tsuda, T. Yokoya, Y. Takano, H. Kito, A. Matsushita, F. Yin, J. Itoh, H. Harima and S. Shin, Phys. Rev. Lett. \textbf{91},127001 (2003).
\bibitem{choi}H. J. Choi, D. Roundy, H. Sun, M. L. Cohen, and S. G. Louie, Nature \textbf{418}, 758 (2002).
\bibitem{guritanu} V. Guritanu, W. Goldacker, F. Bouquet, Y. Wang, R. Lortz, G. Goll, and A. Junod, Phys. Rev. B \textbf{70}, 184526 (2004) 
\bibitem{seyfarth}G. Seyfarth, J. P. Brison, M. A. Measson, J. Flouquet, K. Izawa, Y. Matsuda, H. Sugawara, and H. Sato, Phys. Rev. Lett. \textbf{95}, 107004 (2005).
\bibitem{kasahara} Y. Kasahara, T. Iwasawa, H. Shishido, T. Shibauchi, K. Behnia, Y. Haga, T. D. Matsuda, Y. Onuki, M. Sigrist, and Y. Matsuda, Phys. Rev. Lett. \textbf{99}, 116402 (2007). 
\end{thebibliography}
\end{document}